\documentclass[twocolumn,superscriptaddress,prl]{revtex4-1}
\usepackage{amsmath}
\usepackage{natbib}
\usepackage{graphicx}
\usepackage{cancel}
\begin{document}
\title{Nonuniform currents and spins of relativistic electron vortices in a magnetic field.}
\author{Koen van Kruining}
\email{koen@pks.mpg.de}
\author{Armen G. Hayrapetyan}
\affiliation{Max-Planck-Institut f\"ur Physik komplexer Systeme, 01187 Dresden, Germany}
\author{J\"org B. G\"otte}
\affiliation{Nanjing University, Nanjing 210093, China}
\affiliation{University of Glasgow, Glasgow G12 8QQ, United Kingdom}
\begin{abstract}
We present a relativistic description of electron vortex beams in a homogeneous magnetic field. Including spin from the beginning reveals that spin-polarized electron vortex beams have a complicated azimuthal current structure, containing small rings of counterrotating current between rings of stronger corotating current. Contrary to many other problems in relativistic quantum mechanics, there exists a set of vortex beams with exactly zero spin-orbit mixing in the highly relativistic and nonparaxial regime. The well defined phase structure of these beams is analogous to simpler scalar vortex beams, owing to the protection by the Zeeman effect. For states that do show spin-orbit mixing, the spin polarization across the beam is nonuniform rendering the spin and orbital degrees of freedom inherently inseparable.
\end{abstract}
\maketitle
\textit{Introduction}---
The concept of light beams carrying orbital angular momentum along the propagation axis has been widely utilized in modern optics \cite{Allen+:PiO39:1999, MolinaTerriza+:NatPhys3:2007,FrankeArnold+:LPR2:2008}. Based on analogies of the governing wave equations, vortex beams have also been predicted and generated for electrons \cite{BliokhBliokhSNori07,*BliokhDennisNori11, * BliokhNori12b,UchidaTonomura:Nature464:2010,Verbeeck+:Nature467:2010,McMorran+:Science331:2011,VerbeeckSchatschneiderLS-PLS-TT11,vanBoxemVerbeeckPartoens13,GGMFKB14,*GGMFKB15,*GrilloKarimiGFDennisBoyd14,Laroqueteal16,TTMousleyBabikerY17} and neutrons \cite{Clark+:Nature525:2015}, as well as proposed for atoms \cite{HayrapetyanMatulaSurzhykovFritzsche13,LEBabikerAlD14}. This promises the ability to probe and manipulate matter on smaller length scales, but also opens up the possibility to consider the interaction of vortex beams with external fields \cite{BliokhSchattschneiderVNori12,GreenshieldsStampsFranke-Arnold12, *GreenshieldsFranke-ArnoldStamps15,Karlovets12,HayrapetyanMAielloSFritzsche14, GSchattshneiderBliokhNVerbeeck13, *SchattschneiderSS-PLofflerS-TBliokhNori14}, other vortex beams \cite{Bialynicki-Birula04, *Bialynicki-BirulaChmura05,*Bialynicki-BirulaRadozycki06, Ivanov11, *IvanovSerbo11, *Ivanov12, *IvanofSSurzhykovFritzsche16} and atoms \cite{Matulaetal14, *Serboetal15, *ZaytsevSerboShabajev17}. 

In the simplest description these vortex beams are scalar and obey the paraxial Schr\"odinger equation. Going beyond the paraxial approximation reveals a linking between the spin and orbital degrees of freedom arising whenever the beam is tightly confined, complicating the vortex structure \cite{Barnett17, Bialynicki-Birula17}. And whereas light beams as solutions of Maxwell's equation are naturally relativistic, for particles it is important to distinguish between the nonrelativisitic regime based on Schr\"odinger's equation and the relativistic regime covered by the Dirac equation. 

Whether or not a nonrelativistic description suffices depends not only on the energy of the electron beam involved, but also on the importance the spin of the particle in the interaction in question, as spin is naturally included in the Dirac equation \cite{Dirac28, LL4}. For electrons traveling through a magnetic field it is of particular importance to take the spin into account, because it interacts strongly with the field.


We analytically solve the Dirac equation for an electron in a homogeneous magnetic field, a problem first considered by Landau \cite{LL3,LL4}. The interaction with the magnetic field confines the beam and gives rise to a set of discrete energy levels (Landau levels) \cite{LL3, BliokhSchattschneiderVNori12}. On top of that the Zeeman effect shifts the energy of the positive and negative spin states relative to each other. The quantized Landau and Zeeman contributions to the energy determine which states  undergo spin-orbit mixing with each other and completely forbid spin-orbit mixing for some of them. The inclusion of spin also leads to a (for some states large) redistribution of the azimuthal current within the beam, revealing a pattern of concentric rings of clockwise and counterclockwise rotating current. Our results and conclusions are not only applicable to electrons propagating in beams, but also for electrons confined in Penning traps. 

Throughout this letter we set $c=\hbar=1$, use the standard representation for the Dirac matrices, slashes to denote contraction with Dirac matrices, the positive z-axis as quantization axis for angular momentum and the metric signature diag($+$ $-$ $-$ $-$). 

\begin{figure}
\includegraphics[width=\columnwidth]{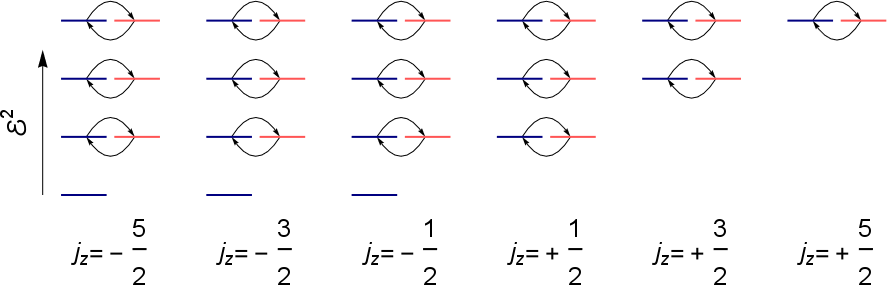}
\caption{The energy levels for a fixed value of $k$ sorted by their total angular momentum. The states with positive spin (red) have one quantum of squared energy more than the states with negative spin (blue). Thus the ground states are not degenerate with any opposite spin states and cannot spin-orbit mix as indicated by the arrows. \label{niveauschema}}
\end{figure}

\textit{Electron beams in a magnetic field and their spin-orbit structure}---
A magnetic field can be incorporated in the Dirac equation using the gauge covariant momentum operator $P_\mu=p_\mu-eA_\mu=i\partial_\mu-eA_\mu$, with $A_\mu$ the vector potential and $e$ the electron charge. Choosing the magnetic field in the positive z-direction we take the vector potential $A_\mu=\frac 12 B(0,-y,x,0)$, with $B$ the magnitude of the magnetic field. Using cylindrical coordinates and first solving the `squared' Dirac equation $(\cancel P+m)(\cancel P-m)\Psi=0$, we assume a solution of the form $\Psi=e^{i(kz-\mathcal E t\pm l\phi)}\psi(r)u$, with $\mathcal E$ the total energy, $u$ a bispinor and $l$ positive. We rescale the radial coordinate $r$ as $\tilde r=\sqrt{|e|B/2}r$.  At a field strength of one Tesla $\tilde r=1$ corresponds to 36 nanometer. The rescaled equation for the spin and radial parts becomes
\begin{multline}
\frac{B|e|}2\left(\frac 1{\tilde r}\partial_{\tilde r}\tilde r\partial_{\tilde r}-\frac{l^2}{\tilde r^2}\mp 2 l-\tilde r^2-2\Sigma_z\right)\psi(\tilde r)u=\\
-(\mathcal E_L^2+\mathcal E_Z^2)\psi(\tilde r)u,\nonumber
\end{multline}
with $\Sigma_i=\mbox{diag}(\hat \sigma_i,\hat \sigma_i)$, and $\hat \sigma_i$ the Pauli matrices. The interaction energy of the electrons spin magnetic moment is $\mathcal E^2_Z=2\sigma_z B|e|$(=Zeeman energy, $\sigma_z=\pm\frac 12$). $\mathcal E_L$ Is the sum of the electrons orbital kinetic energy and the interaction energy of the orbital magnetic moment (Landau energy). The radial differential equation has the well-known solution \cite{LL3, BliokhSchattschneiderVNori12}
$$
\psi(\tilde r)=\tilde r^le^{-\frac{\tilde r^2}2}L_p^l(\tilde r^2),\quad \mathcal E_L^2=B|e|(2p+l(1\pm 1)+1),
$$
with $L_p^l$ an associated Laguerre polynomial. Here the $\pm$-sign is the sign of the orbital angular momentum. For negative orbital angular momentum $\mathcal E_L^2$ is independent of $l$ because the kinetic and magnetic contributions cancel (FIG.~\ref{niveauschema}). These solutions are nondiffracting Laguerre-Gauss beams, with $p$ the radial quantum number indicating how many  rings surround the central spot or ring.  The solutions of the squared Dirac equation describe superpositions of positive and negative energy states. Applying $\cancel P+m$ to the wave functions projects out the positive energy part  (the full calculation is in the supplementary material). The physical solutions are
\begin{widetext}\begin{eqnarray}
&\Psi=&e^{i(kz-\mathcal E t+l\phi)-\frac{\tilde r^2}2}\left(
\tilde r^l L_p^l(\tilde r^2)\left[\begin{array}{c}m+\mathcal E\\0\\k\\0\end{array}\right]+
\sqrt 2i e^{i\phi}\tilde r^{l+1}L^{l+1}_p(\tilde r^2)\left[\begin{array}{c}0\\0\\0\\ \sqrt{B|e|}\end{array}\right]
\right),\qquad \mbox{spin}>0\mbox{, OAM}\ge 0, \nonumber\\
&\Psi=&e^{i(kz-\mathcal E t+l\phi)-\frac{\tilde r^2}2}\left(
\tilde r^l L_p^l(\tilde r^2)\left[\begin{array}{c}0\\m+\mathcal E\\0\\-k\end{array}\right]-
\sqrt 2(p+l)i e^{-i\phi}\tilde r^{l-1}L^{l-1}_p(\tilde r^2)\left[\begin{array}{c}0\\0\\ \sqrt{B|e|}\\0\end{array}\right]
\right),\qquad \mbox{spin}<0\mbox{, OAM}> 0,\nonumber\\
&\Psi=&e^{i(kz-\mathcal E t-l\phi)-\frac{\tilde r^2}2}\left(
\tilde r^l L_p^l(\tilde r^2)\left[\begin{array}{c}m+\mathcal E\\0\\k\\0\end{array}\right]-
\sqrt 2(p+1)i e^{i\phi}\tilde r^{l-1}L^{l-1}_{p+1}(\tilde r^2)\left[\begin{array}{c}0\\0\\0\\ \sqrt{B|e|}\end{array}\right]
\right),\qquad \mbox{spin}>0\mbox{, OAM}< 0,\nonumber\\
&\Psi=&e^{i(kz-\mathcal E t-l\phi)-\frac{\tilde r^2}2}\left(
\tilde r^l L_p^l(\tilde r^2)\left[\begin{array}{c}0\\m+\mathcal E\\0\\-k\end{array}\right]+
\sqrt 2i e^{-i\phi}\tilde r^{l+1}L^{l+1}_{p-1}(\tilde r^2)\left[\begin{array}{c}0\\0\\ \sqrt{B|e|}\\0\end{array}\right]
\right),\qquad \mbox{spin}<0\mbox{, OAM}\le 0,\nonumber
\end{eqnarray}\end{widetext}
for each of the four combinations of positive and negative spin and orbital angular momentum. Whenever we derive an expression which is different for these four solutions, we put the corresponding expressions in the same order. The second term in the brackets is the spin-orbit mixing term, which appears because orbital angular momentum is not a good quantum number \cite{Dirac28}. 

Of particular interest is the last expression (negative spin and orbital angular momentum). Rewriting $L_{p-1}^{l+1}=-{L'}_p^l$ \cite{RHB}, one sees that the spin-orbit term is zero for $p=0$. The lack of spin-orbit mixing for these states stems from all states having a well defined angular momentum and squared energy. The Zeeman effect shifts the squared energy upwards by $\mathcal E_Z^2=B|e|$ for the states with positive spin and downwards by the same amount for the states with negative spin. The Landau quantization generates a squared energy ladder with level spacing $\Delta\mathcal E_L^2=2B|e|$, twice the Zeeman shift. So the positive spin states are shifted upward one level compared to the negative spin states (FIG.~\ref{niveauschema}) and for the lowest lying states with negative spin there is no positive spin state with equal squared energy they can spin-orbit mix with.
Without the spin-orbit mixing term, the wave function factorizes into a product state of a constant bispinor and a scalar function. Typically, both for light and electrons, such a simple separation in a spin part and a spatial part is not possible, making these negative angular momentum $p=0$ states quite special. This clean separation of spin and orbital angular momentum also makes the ground states perfectly spin polarized, a condition which otherwise has only been achieved with a more complicated combination of magnetic and electric fields \cite{Karimietal12, Nsofinietal16} high loss of beam intensity \cite{SchattschneiderGrilloAubry17} or extremely high laser intensities \cite{DellwegMuller17}. That they are (for a given $k$) the lowest energy states suggests that there should be a way to selectively populate these `scalar like' unperturbed nonparaxial vortex states. 

\textit{Detailed analysis of the current structure}---
The detailed charge flow within the beam can be computed using the four current $j_\mu=\Psi^\dagger\gamma_0\gamma_\mu\Psi$. Integrating its zeroth component over the entire transverse plane gives a  useful normalization factor. Using $\int_0^\infty x^l L_p^l(x)^2e^{-x}dx=\frac{(l+p)!}{p!}$, the integrated probability density is evaluated to be resp.
\begin{eqnarray}
&\textstyle{\int} j_0=& \pi{\textstyle\frac{(l+p)!}{p!}}\left(m^2+\mathcal E^2+2m\mathcal E+k^2+2B|e|(l+p+1)\right),\nonumber\\
&\textstyle{\int} j_0=& \pi{\textstyle\frac{(l+p)!}{p!}}\left(m^2+\mathcal E^2+2m\mathcal E+k^2+2B|e|(l+p)\right),\nonumber\\
&\textstyle{\int} j_0=& \pi{\textstyle\frac{(l+p)!}{p!}}\left(m^2+\mathcal E^2+2m\mathcal E+k^2+2B|e|(p+1)\right),\nonumber\\
&\textstyle{\int} j_0=& \pi{\textstyle\frac{(l+p)!}{p!}}\left(m^2+\mathcal E^2+2m\mathcal E+k^2+2B|e|p\right).\nonumber
\end{eqnarray}
The last term in the brackets is in each case $\mathcal E_L^2+\mathcal E_Z^2$. Using $\mathcal E=\sqrt{m^2+k^2+\mathcal E_L^2+\mathcal E_Z^2}$, the integrated probability density can in each case be written as ${\int }j_0=2\pi \mathcal E(\mathcal E+m)\frac{(l+p)!}{p!}$. The total current in the z-direction through the transverse plane is
$$
\textstyle{\int }j_z={ \int} j_0\displaystyle \frac k{\mathcal E},
$$
so the electrons have the same speed as particles with mass $\sqrt{m^2+\mathcal E_L^2+\mathcal E_Z^2}$. For the transverse current components one can transform the Dirac matrices into 
\begin{eqnarray}
&\gamma_r=&\cos\phi\gamma_x+\sin\phi\gamma_y,\nonumber\\
&\gamma_\phi=&-\sin\phi\gamma_x+\cos\phi\gamma_y.\nonumber
\end{eqnarray}
The radial component is always zero and the azimuthal component is
\begin{eqnarray}
&j_\phi=&2\sqrt 2\tilde r^{2l+1}e^{-\tilde r^2}L_p^l(\tilde r^2)L_p^{l+1}(\tilde r^2)(\mathcal E+m)\sqrt{B|e|},\nonumber\\
&j_\phi=&2\sqrt 2(p+l)\tilde r^{2l-1}e^{-\tilde r^2}L_p^l(\tilde r^2)L_p^{l-1}(\tilde r^2)(\mathcal E+m)\sqrt{B|e|},\nonumber\\
&j_\phi=-&2\sqrt 2(p+1)\tilde r^{2l-1}e^{-\tilde r^2}L_p^l(\tilde r^2)L_{p+1}^{l-1}(\tilde r^2)(\mathcal E+m)\sqrt{B|e|},\nonumber\\
&j_\phi=-&2\sqrt 2\tilde r^{2l+1}e^{-\tilde r^2}L_p^l(\tilde r^2)L_{p-1}^{l+1}(\tilde r^2)(\mathcal E+m)\sqrt{B|e|},\nonumber
\end{eqnarray}
\begin{figure}[h!]
\includegraphics[width=\columnwidth]{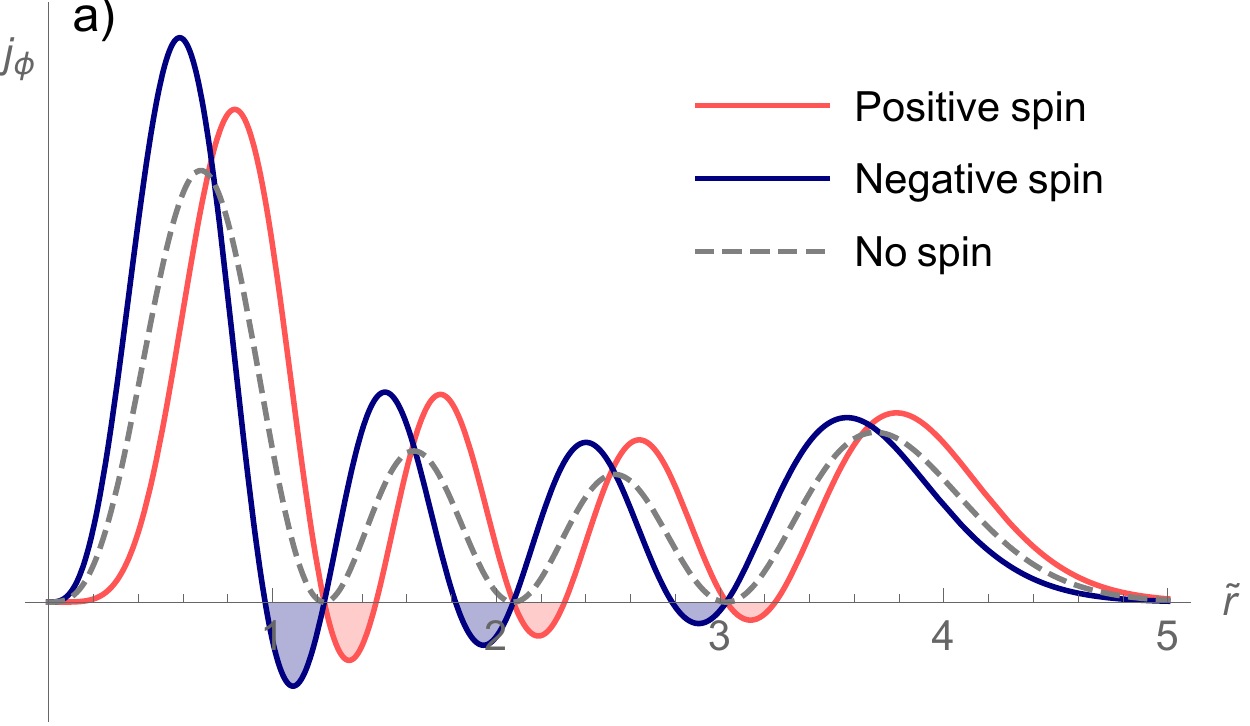}
\includegraphics[width=\columnwidth]{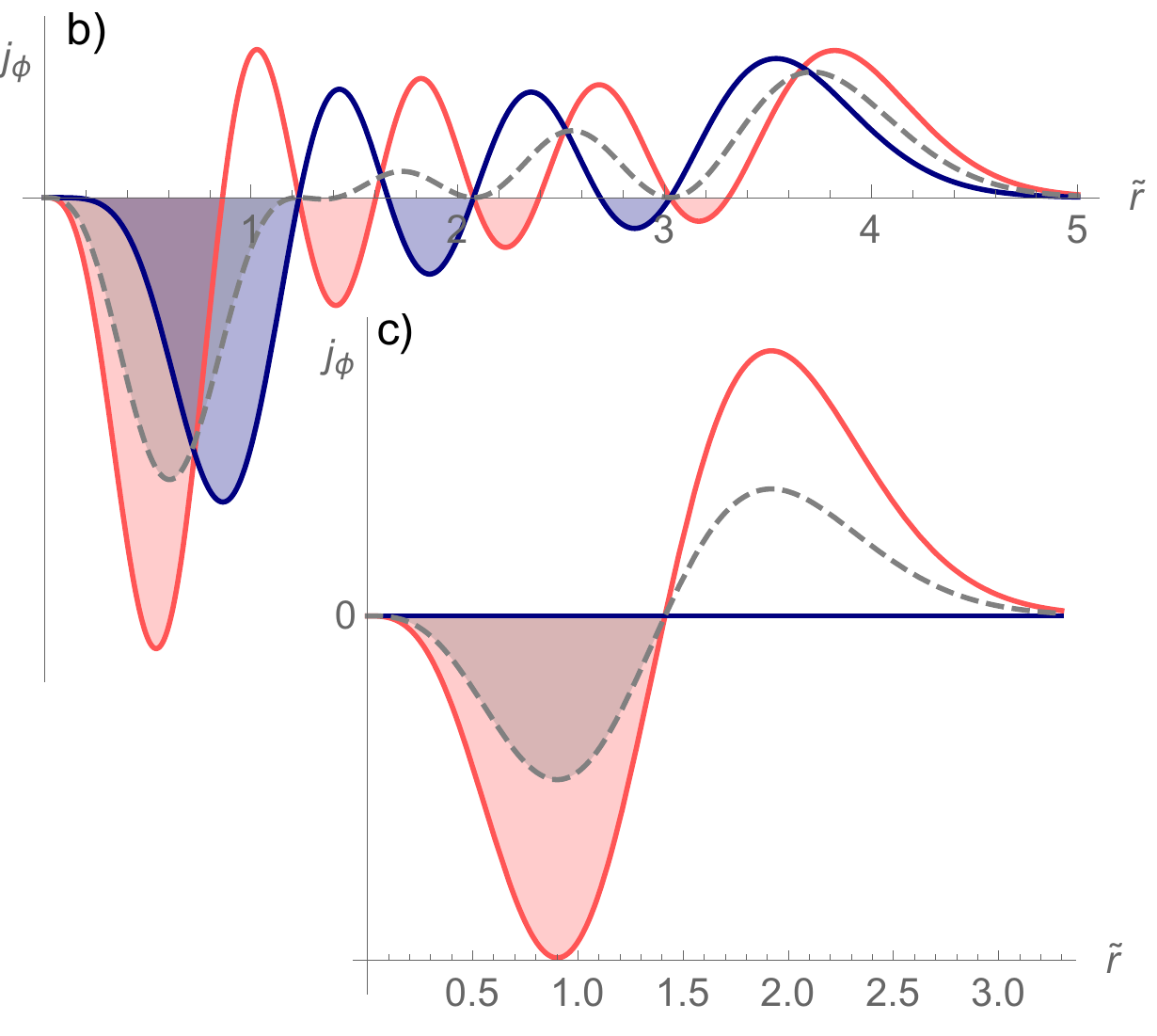}
\caption{$j_\phi$ For positive spin (red), negative spin (blue) and a spin 0 beam for comparison (dashed) for $l=2,p=3$ (a), $l=-2,p=3$ (b) and $l=-2,p=0$ (c). The spin part of the current gives rise to a series of dips where the current flows in the opposite direction, which are absent when spin is neglected. For negative $l$, the azimuthal current is negative near the center but positive on the outside due to the interaction with the magnetic field. The most striking difference from a spin 0 vortex beam occurs for negative $l$ and $p=0$ where negative spin is a Landau-Zeeman ground state and the azimuthal current is exactly zero everywhere.\label{azc}}
\end{figure}
where we have used the surface element $dz\times d\tilde r$. Rescaling $d\tilde r$ back to $dr$ gives a current proportional to $B|e|$. These expressions are quite different from the azimuthal currents for scalar vortex beams in a magnetic field \cite{BliokhSchattschneiderVNori12}, because the spin contribution is included in them as well \cite{Ohanian85}. As can be seen in FIG.~\ref{azc} the inclusion of the spin current reveals complicated patterns of flows and counterflows, which are absent if spin is neglected. These keep their shape even for magnetic field strengths at which there is no appreciable spin-orbit induced change in the beam profile (FIG.~\ref{fig3})
\begin{figure}
\begin{minipage}{0.5\columnwidth}
\includegraphics[width=\columnwidth]{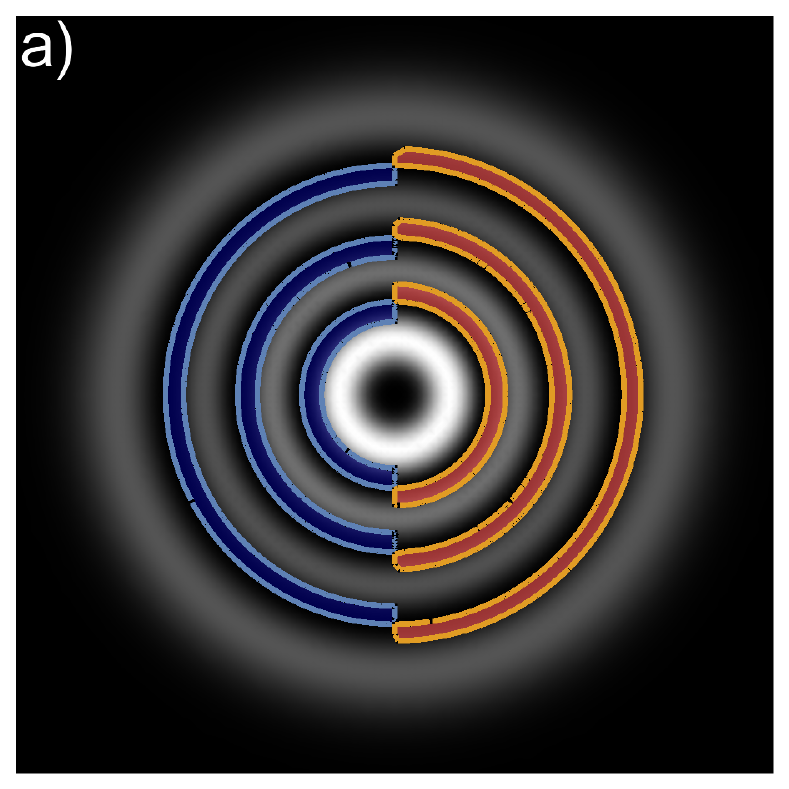}
\end{minipage}\begin{minipage}{0.5\columnwidth}
\includegraphics[width=\columnwidth]{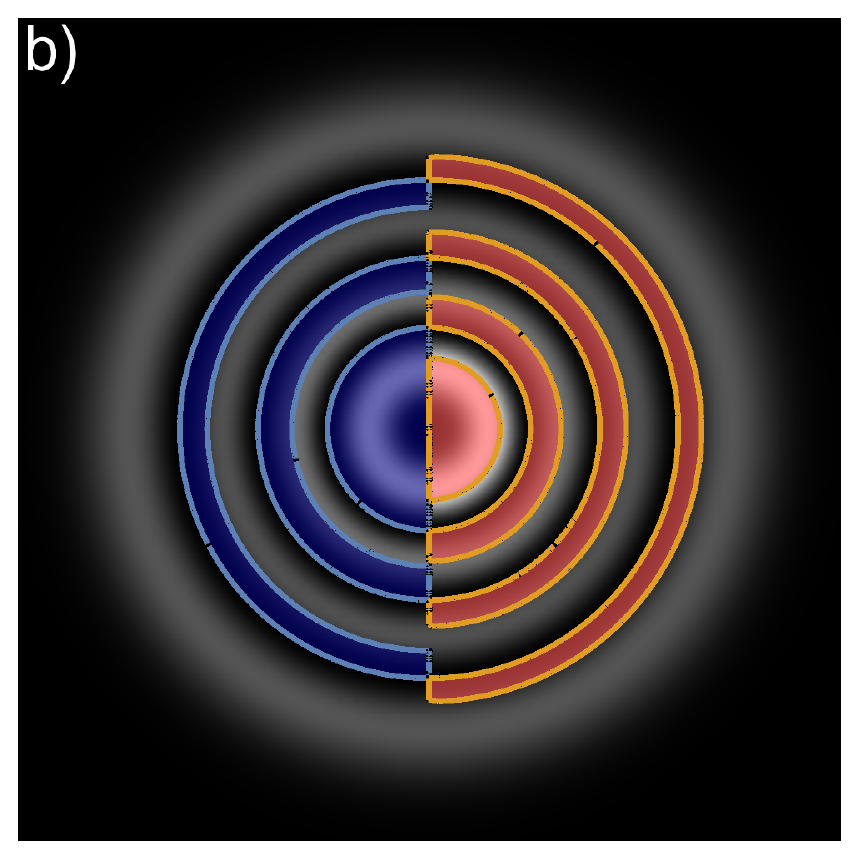}
\end{minipage}
\caption{The regions of negative(=clockwise) azimuthal electron current marked for the $p=3$, $l=2$ state (a) and $p=3$, $l=-2$ (b) for negative (left side, in blue) and positive spin (right side, in red) superposed on the beam profiles for a magnetic field of 1 Tesla. The negative currents occur on the inner side of the dark fringes for positive spin and on the outer side for negative spin. Visible rearrangement of the electron density due to spin-orbit mixing only appears around 1 gigatesla.} \label{fig3}
\end{figure}

\textit{Nonuniform spin}---
As a consequence of spin-orbit mixing, the spin polarization of an electron becomes nonuniform, similar to the nonuniform spin appearing in structured light \cite{BliokhNori12, *BliokhBekshaevNori14, *BekshaevBliokhNori15,NBAB15,Antognozzietal16}, which is used for direction sensitive optical switching \cite{KienHakuta08, *KienRauschenbeutel14a, PhysRevLett.110.213604, *Petersen67,*MSASRauschenbeutel14, *PhysRevX.5.041036, PhysRevLett.113.237203,*PhysRevA.91.042116,*PhysRevA.93.062104, *PhysRevA.93.063830, Kopenhagen15}. Its existence can be inferred decomposing the probability current in a spin and an orbital part \cite{Gordon1928, Ohanian85} and comparing the z-components of the orbital part and the total current, finding that $\mathrm{Re}(\Psi^*P_z/m\Psi)\neq\Psi^*\gamma_0\gamma_z\Psi$. The difference has to be made up for by a spin current $\nabla\times \vec{ \mathcal S}$ caused by a spin component perpendicular to $\hat z$ \footnote{Both Gordon and Ohanian only consider the free case, but because the probability current is gauge invariant, one can see that one must use the gauge covariant momentum operator if fields are present.}. Using 
$\Sigma_r=\cos\phi\Sigma_x+\sin\phi\Sigma_y\mbox{, }
\Sigma_\phi=-\sin\phi\Sigma_x+\cos\phi\Sigma_y,$ 
it can be shown that the radial spin is zero and the azimuthal spin is $\pm\frac 12 k/(\mathcal E+m)j_\phi$, where the sign is given by the sign of the total spin in the z-direction. The ground states' spin polarization is uniform because their spin-orbit mixing is zero. This in contrast to structured light, where the nonuniformity inevitably appears in any finite width beam. 

The difference between a uniformly and a nonuniformly spin polarized state is that for a uniformly polarized state one can always choose a direction along which a spin measurement will certainly give the outcome spin up whereas this is impossible for a nonuniformly polarized state, because spin and spatial degrees of freedom are entangled. For our electron beams this entanglement can be shown by taking their density matrices and tracing out everything except the spin. The remaining mixed spin state is for positive spin
$$
\rho_s=\frac{((m+\mathcal E)^2+k^2)|\uparrow\rangle\langle\uparrow|+(\mathcal E_L^2+\mathcal E_Z^2)|\downarrow\rangle\langle\downarrow|}{2\mathcal E(\mathcal E+m)}
$$
and the same with the spins interchanged for negative spin, showing that one cannot separate the spin and orbital degrees of freedom.

\textit{Gauge covariant angular momentum operator}---
With our choice of gauge, the exact solutions of the Dirac equation are eigenfunctions of the canonical angular momentum operator ($\hat J_z=-i\partial_\phi+\frac 12\Sigma_z$) with eigenvalues resp. $l+\frac 12,\;l-\frac 12,\;-l+\frac 12$ and $-l-\frac 12$. The canonical momentum is not gauge covariant but can be made so by the usual minimal substitution, yielding: $\mathcal {\hat J}_z=-i\partial_\phi-erA_\phi+\frac 12\Sigma_z$. This operator does not have any stationary solution of the Dirac equation or the `squared' Dirac equation as its eigenstate, as can be verified by applying it to any (linear combination of degenerate) basis state. Its expectation value can be computed by adding $\displaystyle\frac{\int\Psi^\dagger |e|rA_\phi\Psi}{\int \Psi^\dagger\Psi}=\displaystyle\frac{\int\Psi^\dagger\tilde r^2\Psi}{\int \Psi^\dagger\Psi}$ to the canonical angular momentum, the result is (suppl. mat.)
\begin{eqnarray}
&\mathcal J_z=&2p+2l+\frac 32+\frac{\mathcal E^2_L+\mathcal E^2_Z}{2\mathcal E(\mathcal E+m)},\nonumber\\
&\mathcal J_z=&2p+2l+\frac 12-\frac{\mathcal E^2_L+\mathcal E^2_Z}{2\mathcal E(\mathcal E+m)},\nonumber \end{eqnarray}\begin{eqnarray}
&\mathcal J_z=&2p+\frac 32+\frac{\mathcal E^2_L+\mathcal E^2_Z}{2\mathcal E(\mathcal E+m)},\nonumber\\
&\mathcal J_z=&2p+\frac 12-\frac{\mathcal E^2_L+\mathcal E^2_Z}{2\mathcal E(\mathcal E+m)}.\nonumber
\end{eqnarray}
If one would neglect the spin-orbit term, one would always get a half-integer expectation value for the gauge covariant angular momentum, although the states, even without spin-orbit term are \emph{not} eigenstates of $\mathcal {\hat J}_z$. This fortuitous coincidence has been overlooked in the literature until now, to the best of our knowledge. The reason that the expectation value of $\mathcal{\hat J}_z$ is not a half integer number is that the orbital contribution changes by two quanta when $l$ or $p$ is changed by one whereas the spin contribution changes by the usual one quantum upon spin flip. Therefore the main term and the spin orbit term have different expectation values for $\mathcal{\hat J}_z$ and one takes the probability weighted average of the both terms. $\mathcal{\hat L}_z+2\mathcal{\hat S}_z$ Does have half-integer expectation values. This last quantity determines the z-component of the magnetic moment, $M_z$, of the electron as can be verified by computing (details in suppl. mat.)
\begin{multline}
M_z=\int \frac e2 rj_\phi=-\frac{\textstyle{\int}j_0}{\mathcal E}\frac{\mathcal E_L^2+\mathcal E_Z^2}{2B}=\\
\frac{e}{2\mathcal E}{\textstyle\int}j_0(2p+l(1\pm1)+1+2\mathcal S)=\frac e{2\mathcal E}\textstyle{\int}j_0(\mathcal L+2\mathcal S)\nonumber
\end{multline}
showing that the gauge covariant operators are the ones determining the magnetic moment.

Apart from not having any stationary eigenfunctions, the gauge covariant angular momentum operators also do not generate a Lie group. These two properties can be proven more generally. Taking the commutator of two of these operators gives (suppl. mat.)
$$
[\mathcal{\hat J}_j,\mathcal{\hat J}_k]=-i\epsilon_{jkl}(\mathcal{\hat J}_l-x_l\mathbf{x\cdot B}),
$$
showing that they violate the closure axiom for Lie algebras if there is any magnetic field present. For the existence of stationary solutions we change notation and write the components of the gauge covariant momenta and `boost' operators as an antisymmetric tensor $\mathcal{\hat J}_{\mu\nu}=x_{[\mu} P_{\nu]}+\frac i2\sigma_{\mu\nu}$. The brackets on the indices indicate antisymmetrization, $T_{[\mu\nu]}=T_{\mu\nu}-T_{\nu\mu}$. With this notation, $\mathcal{\hat J}_{12}=\mathcal{\hat J}_z$. Now the existence of physical states that are eigenstates of $\mathcal{\hat J}_{\mu\nu}$ is only possible if the commutator $[\cancel P-m,\mathcal{\hat J}_{\mu\nu}]$ vanishes. This commutator is (suppl. mat.)
$$
[\cancel P-m,\mathcal{\hat J}_{\mu\nu}]=iex_{[\mu}F_{\nu]\lambda}\gamma^{\lambda},
$$
which vanishes only for an extremely restricted class of possible fields. Taking $\mathcal{\hat J}_{12}$ and writing out the field components explicitly, we have
$$
[\cancel P-m,\mathcal{\hat J}_{12}]=ie\left((xE_y-yE_x)\gamma_0-B_z \mathbf x\cdot\vec\gamma+\gamma_3 \mathbf x\cdot \mathbf B\right).
$$
So the only possible field that would allow for physical eigenstates of $\mathcal{\hat J}_{12}$, is a constant electric field in the z-direction.

\textit{Conclusion}---
We have shown that in a homogeneous magnetic field there exist electron vortex beams without spin-orbit mixing and thus with a very `clean' vortex core. For these beams, spin-orbit mixing remains absent even for strong magnetic fields and relativistic speeds. Including the effect of spin reveals an internal rearrangement of the azimuthal current which is quite substantial if the orbital angular momentum and magnetic field point in opposite directions. For electron vortex beams the current scales linearly with the beam intensity and the spin rearrangement of the azimuthal current can be magnified by using a strong enough electron beam. If an electron vortex beam is wide enough, a suitable test particle can probe these current rearrangements similar to how a small dielectric particle can probe the local Poynting vector of a light beam \cite{HeFrieseHeckenbergRubinsztein-Dunlop95, PhysRevLett.88.053601,NBAB15,Antognozzietal16}. 

\textit{Acknowledgements}--- KvK thanks Valentin Walther for the useful discussion helping him clarify the nature of the nonuniform electron spin. This work was supported by the Engineering and Physical Sciences Research Council of the United Kingdom with grants Nos. EP/I012451/1 and EP/M01326X/1 and the the National Key Research and Development Program of China under contract number 2017YFA0303700.
\bibliography{/home/koen/Documents/Publicaties/Landaubundels/Landaubeams.bib}

\onecolumngrid
\pagebreak
\begin{center}
\Large Supplementary material
\end{center}
\subsection*{Obtaining the exact solutions of the Dirac equation in a magnetic field}
The solutions of te squared Dirac equation in a constant magnetic field (symmetric gauge) are, using the rescaled coordinate $\tilde r=\sqrt{\frac{|e|B}2}r$ and taking $l$ positive
$$\Psi=e^{i(kz-\mathcal E t\pm l\phi)}\tilde r^{l}e^{-\frac{\tilde r^2}2}L_p^{l}(\tilde r^2)\left(\left[\begin{array}{c}1\\0\\0\\0\end{array}\right]\vee\left[\begin{array}{c}0\\1\\0\\0\end{array}\right]\right).$$
The exact solutions pf the first order Dirac equation can be obtained by applying $\cancel P+m$ to the solutions of the squared Dirac equation. Using
$$
(\cancel P+m)\left[\begin{array}{c}1\\0\\0\\0\end{array}\right]=\left[\begin{array}{c}
i\partial_t+m\\0\\-i\partial_z\\ -i\partial_x+\partial_y+\frac{|e|B}2(i x-y)
\end{array}\right],\qquad (\cancel P+m)\left[\begin{array}{c}0\\1\\0\\0\end{array}\right]=\left[\begin{array}{c}
0\\ i\partial_t+m\\-i\partial_x-\partial_y-\frac{|e|B}2(ix+y)\\-i\partial_z
\end{array}\right],
$$
the de derivatives with respect to $t$ and $z$ are easy to compute and give resp. $\mathcal E$ and $k$. For the transverse derivatives, one can use the rescaled coordinates $\tilde x=\sqrt{\frac{|e|B}2}x,\; \tilde y=\sqrt{\frac{|e|B}2}y$ to rewrite them as
$$
(\cancel P+m)\left[\begin{array}{c}1\\0\\0\\0\end{array}\right]=\left[\begin{array}{c}
m+\mathcal E\\0\\k\\ \sqrt{\frac{|e|B}2}(-i\partial_{\tilde x}+\partial_{\tilde y}+(i \tilde x-\tilde y))
\end{array}\right],\qquad (\cancel P+m)\left[\begin{array}{c}0\\1\\0\\0\end{array}\right]=\left[\begin{array}{c}
0\\m+\mathcal E\\\sqrt{\frac{|e|B}2}(-i\partial_{\tilde x}-\partial_{\tilde y}-(i\tilde x+\tilde y))\\k
\end{array}\right].
$$
The components $\sqrt{\frac{|e|B}2}(-i\partial_{\tilde x}+\partial_{\tilde y}+(i \tilde x-\tilde y))$ and $\sqrt{\frac{|e|B}2}(-i\partial_{\tilde x}-\partial_{\tilde y}-(i\tilde x+\tilde y))$ give rise to the spin-orbit mixing terms, whose explicit computation is rather lengthy. The following three identities will be of use
\begin{eqnarray}
(\partial_{\tilde x}\pm i\partial_{\tilde y})\tilde r^n&=&n\tilde r^{n-1}e^{\pm i\phi},\nonumber\\
(\partial_{\tilde x}\pm\partial_{\tilde y})\tilde r^{|n|}e^{\pm i|n|\phi}&=&0,\nonumber\\
(\partial_{\tilde x}\pm\partial_{\tilde y})\tilde r^{|n|}e^{\mp i|n|\phi}&=&2|n|\tilde r^{|n|-1}e^{\pm i(|n|-1)\phi}.\nonumber
\end{eqnarray}
The form of the spin-orbit term depends on the signs of the spin and orbital angular momentum. For spin and orbital angular momentum positive, one has
\begin{multline}
 \sqrt{\frac{|e|B}2}(-i\partial_{\tilde x}+\partial_{\tilde y}+(i \tilde x-\tilde y))e^{i(kz-\mathcal E t+ l\phi)}\tilde r^{l}e^{-\frac{\tilde r^2}2}L_p^{l}(\tilde r^2)=\\
i \sqrt{2|e|B}\left(e^{i(kz-\mathcal E t+ (l+1)\phi)}\tilde r^{l+1}e^{-\frac{\tilde r^2}2}\left(L_p^l(\tilde r^2)-{L'}_p^{l}(\tilde r^2)\right)\right).\nonumber
\end{multline}
Using the recurrence relations for Laguerre polynomials ${L'}_p^l(\tilde r^2=-L_{p-1}^{l+1}(\tilde r^2)$ (prime denotes differentiation with respect to $\tilde r^2$) and $L_p^l(\tilde r^2)=L_p^{l+1}(\tilde r^2)-L_{p-1}^{l+1}(\tilde r^2)$, the Laguerre polynomials in the brackets become simply $L_p^{l+1}(\tilde r^2)$, thus the overall solution of the first order Dirac equation in this case becomes
$$
\Psi=e^{i(kz-\mathcal E t+l\phi)-\frac{\tilde r^2}2}\left(
\tilde r^l L_p^l(\tilde r^2)\left[\begin{array}{c}m+\mathcal E\\0\\k\\0\end{array}\right]+
\sqrt 2i e^{i\phi}\tilde r^{l+1}L^{l+1}_p(\tilde r^2)\left[\begin{array}{c}0\\0\\0\\ \sqrt{B|e|}\end{array}\right]
\right). \nonumber\\
$$

Now for positive orbital angular momentum and negative spin, the spin-orbit term is
\begin{multline}
 \sqrt{\frac{|e|B}2}(-i\partial_{\tilde x}-\partial_{\tilde y}-(i \tilde x+\tilde y))e^{i(kz-\mathcal E t+ l\phi)}\tilde r^{l}e^{-\frac{\tilde r^2}2}L_p^{l}(\tilde r^2)=\\
i \sqrt{2|e|B}\left(e^{i(kz-\mathcal E t+ (l-1)\phi)}\tilde r^{l-1}e^{-\frac{\tilde r^2}2}\left(-lL_p^l(\tilde r^2)-\tilde r^2{L'}_p^{l}(\tilde r^2)\right)\right).\nonumber
\end{multline}
Using the recurrence relation
$$
\tilde r^2 {L'}_p^{l}(\tilde r^2)=pL_p^{l}(\tilde r^2)-(p+l)L_{p-1}^{l}(\tilde r^2)
$$
to rewrite the derivatives of the Laguerre polynomial, one gets
$$
-\left(lL_p^l(\tilde r^2)+pL_p^l(\tilde r^2)-(p+l)L_{p-1}^l(\tilde r^2)\right).
$$
With the relation $L_p^{l-1}(\tilde r^2)=L_p^l(\tilde r^2)-L_{p-1}^l(\tilde r^2)$ this expression simplifies to $-(p+l)L_{p}^{l-1}(\tilde r^2)$ and the solution of the first order Dirac equation for positive orbital angular momentum and negative spin becomes
$$
\Psi=e^{i(kz-\mathcal E t+l\phi)-\frac{\tilde r^2}2}\left(
\tilde r^l L_p^l(\tilde r^2)\left[\begin{array}{c}0\\m+\mathcal E\\0\\-k\end{array}\right]-
\sqrt 2(p+l)i e^{-i\phi}\tilde r^{l-1}L^{l-1}_p(\tilde r^2)\left[\begin{array}{c}0\\0\\ \sqrt{B|e|}\\0\end{array}\right]
\right).
$$
For negative orbital angular momentum and positive spin, one has
\begin{multline}
 \sqrt{\frac{|e|B}2}(-i\partial_{\tilde x}+\partial_{\tilde y}+(i \tilde x-\tilde y))e^{i(kz-\mathcal E t-l\phi)}\tilde r^{l}e^{-\frac{\tilde r^2}2}L_p^{l}(\tilde r^2)=\\
i \sqrt{2|e|B}\left(e^{i(kz-\mathcal E t+ (l+1)\phi)}\tilde r^{l-1}e^{-\frac{\tilde r^2}2}\left(\tilde r^2 L_p^l(\tilde r^2)-lL_p^l(\tilde r^2)-\tilde r^2{L'}_p^{l}(\tilde r^2)\right)\right).\nonumber
\end{multline}
Using $L_p^l(\tilde r^2)=-{L'}_{p-1}^{l+1}(\tilde r^2)$, where the prime denotes differentiation with respect to $\tilde r^2$, one can rewrite the first Laguerre polynomial:
$$
-\left(r^2 {L'}^{l-1}_{p+1}(\tilde r^2)+lL_p^l(\tilde r^2)+\tilde r^2{L'}_p^{l}(\tilde r^2)\right).
$$
Using again $\tilde r^2 {L'}_p^{l}(\tilde r^2)=pL_p^{l}(\tilde r^2)-(p+l)L_{p-1}^{l}(\tilde r^2)$ this expression becomes
$$
-\left((p+1)L_{p+1}^{l-1}(\tilde r^2)-(p+l)L_p^{l-1}(\tilde r^2)+lL_p^l(\tilde r^2)+pL_p^l(\tilde r^2)-(p+l)L_{p-1}^l(\tilde r^2)\right).
$$
Because of $L_p^{l-1}(\tilde r^2)=L_p^l(\tilde r^2)-L_{p-1}^l(\tilde r^2)$, everything but the first term cancels and the solution for the first order Dirac equation for negative orbital angular momentum and positive spin becomes
$$
\Psi=e^{i(kz-\mathcal E t-l\phi)-\frac{\tilde r^2}2}\left(
\tilde r^l L_p^l(\tilde r^2)\left[\begin{array}{c}m+\mathcal E\\0\\k\\0\end{array}\right]-
\sqrt 2(p+1)i e^{i\phi}\tilde r^{l-1}L^{l-1}_{p+1}(\tilde r^2)\left[\begin{array}{c}0\\0\\0\\ \sqrt{B|e|}\end{array}\right]
\right)
$$
The case of negative orbital angular momentum and spin is simple, one has
$$
\sqrt{\frac{|e|B}2}(-i\partial_{\tilde x}-\partial_{\tilde y}-(i \tilde x+\tilde y))e^{i(kz-\mathcal E t+ l\phi)}\tilde r^{l}e^{-\frac{\tilde r^2}2}L_p^{l}(\tilde r^2)=
-i \sqrt{2|e|B}\left(e^{i(kz-\mathcal E t+ (l-1)\phi)}\tilde r^{l-1}e^{-\frac{\tilde r^2}2}\tilde r^2{L'}_p^{l}(\tilde r^2)\right),\nonumber
$$
and using again $L_p^l(\tilde r^2)=-{L'}_{p-1}^{l+1}(\tilde r^2)$, the overall solution of the first order Dirac equation becomes
$$
\Psi=e^{i(kz-\mathcal E t-l\phi)-\frac{\tilde r^2}2}\left(
\tilde r^l L_p^l(\tilde r^2)\left[\begin{array}{c}0\\m+\mathcal E\\0\\-k\end{array}\right]+
\sqrt 2i e^{-i\phi}\tilde r^{l+1}L^{l+1}_{p-1}(\tilde r^2)\left[\begin{array}{c}0\\0\\ \sqrt{B|e|}\\0\end{array}\right]
\right)
$$
\subsection*{Explicit forms of the radial and azimuthal gamma and spin matrices}
\begin{eqnarray}
\gamma^r=\cos \phi \gamma^x+\sin \phi \gamma^y=\left[\begin{array}{c c c c}
0 & 0 & 0& e^{-i\phi}\\ 0& 0 & e^{i\phi}& 0\\ 0 & -e^{-i\phi} & 0 & 0\\ -e^{i\phi}& 0 & 0 & 0
\end{array}\right],\nonumber\\
\gamma^\phi=-\sin\phi \gamma^x+\cos\phi\gamma^y=\left[\begin{array}{c c c c}
0 & 0 & 0& -ie^{-i\phi}\\ 0& 0 & ie^{i\phi}& 0\\ 0 & ie^{-i\phi} & 0 & 0\\ -ie^{i\phi}& 0 & 0 & 0
\end{array}\right],\nonumber\\
\Sigma_r=\cos\phi\Sigma_x+\sin\phi\Sigma_y=\left[\begin{array}{c c c c}
0&e^{-i\phi} &0&0\\ e^{i\phi} &0 & 0&0\\ 0&0&0&e^{-i\phi}\\ 0&0&e^{i\phi}&0
\end{array}\right],\nonumber\\
\Sigma_\phi=-\sin\phi\Sigma_x+\cos\phi\Sigma_y=\left[\begin{array}{c c c c}
0& -i e^{-i\phi}&0&0\\ ie^{i\phi}& 0 & 0&0\\ 0&0&0& -ie^{-i\phi}\\ 0&0&ie^{i\phi}&0
\end{array}\right].\nonumber
\end{eqnarray}
\subsection*{Explicit evaluation of $\displaystyle\frac{\int\Psi^\dagger\tilde r^2\Psi}{\int \Psi^\dagger\Psi}$.}
The quantity $\displaystyle\frac{\int\Psi^\dagger\tilde r^2\Psi}{\int \Psi^\dagger\Psi}$ appears in the computation of the gauge covariant angular momentum of the electron vortex states. For the four different combinations of positive and negative orbital angular momentum and spin (order the same as in the main text)  $\int\Psi^\dagger\tilde r^2\Psi$ can be shown to be resp.
\begin{eqnarray}
\int\Psi^\dagger\tilde r^2\Psi=2\pi\int_0^\infty\left((\mathcal E+m)^2+k^2\right)\tilde r^{2l+2}{L^l_p}^2(\tilde r^2) e^{-\tilde r^2}+2 B|e| \tilde r^{2l+4}{L_p^{l+1}}^2(\tilde r^2) e^{-\tilde r^2} d\tilde r,\nonumber\\
\int\Psi^\dagger\tilde r^2\Psi=2\pi\int_0^\infty\left((\mathcal E+m)^2+k^2\right)\tilde r^{2l+2}{L^l_p}^2(\tilde r^2) e^{-\tilde r^2}+2B|e|(p+l)^2\tilde r^{2l}{L_p^{l-1}}^2(\tilde r^2)e^{-\tilde r^2}d\tilde r,\nonumber\\
\int\Psi^\dagger\tilde r^2\Psi=2\pi\int_0^\infty\left((\mathcal E+m)^2+k^2\right)\tilde r^{2l+2}{L^l_p}^2(\tilde r^2) e^{-\tilde r^2}+2B|e|(p+1)^2\tilde r^{2l}{L_{p+1}^{l-1}}^2(\tilde r^2)e^{-\tilde r^2}d\tilde r\nonumber\\
\int\Psi^\dagger\tilde r^2\Psi=2\pi\int_0^\infty\left((\mathcal E+m)^2+k^2\right)\tilde r^{2l+2}{L^l_p}^2(\tilde r^2) e^{-\tilde r^2}+2B|e|\tilde r^{2l+4}{L_{p-1}^{l+1}}^2(\tilde r^2)e^{-\tilde r^2}d\tilde r.\nonumber
\end{eqnarray}
Substituting $x=\tilde r^2$ turns these integrals to
\begin{eqnarray}
\pi\int_0^\infty\left((\mathcal E+m)^2+k^2\right)x^{l+1}{L^l_p}^2(x) e^{-x}+2 B|e| x^{l+2}{L_p^{l+1}}^2(x) e^{-x} dx,\nonumber\\
\pi\int_0^\infty\left((\mathcal E+m)^2+k^2\right)x^{l+1}{L^l_p}^2(x) e^{-x}+2B|e|(p+l)^2x^l{L_p^{l-1}}^2(x)e^{-x}dx,\nonumber\\
\pi\int_0^\infty\left((\mathcal E+m)^2+k^2\right)x^{l+1}{L^l_p}^2(x) e^{-x}+2B|e|(p+1)^2x^l{L_{p+1}^{l-1}}^2(x)e^{-x}dx\nonumber\\
\pi\int_0^\infty\left((\mathcal E+m)^2+k^2\right)x^{l+1}{L^l_p}^2(x) e^{-x}+2B|e|x^{l+2}{L_{p-1}^{l+1}}^2(x)e^{-x}dx.\nonumber
\end{eqnarray}
For Laguerre polynomials, we have the orthogonality relation $\int_0^\infty L_p^l(x)L_{p'}^l(x)e^{-x}dx=\frac{(l+p)!}{p!}\delta_{pp'}$. Using this relation and $L_p^{l-1}(x)=L_p^l(x)-L_{p-1}^l(x)$, we obtain the following integral identity
$$
\int_0^\infty x^{l+1}{L_p^l}^2(x)e^{-x}dx=\int_0^\infty x^{l+1}\left(L_p^{l+1}(x)-L_{p-1}^{l+1}(x)\right)^2e^{-x}dx=\frac{(l+p+1)!}{p!}+\frac{l+p!}{(p-1)!}=\frac{l+p!}{p!}(2p+l+1),
$$
which can be used to evaluate all the integals and obtain
\begin{eqnarray}
\int\Psi^\dagger\tilde r^2\Psi=\pi\frac{(l+p)!}{p!}\left(\left((\mathcal E+m)^2+k^2\right)(2p+l+1)+2B|e|(p+l+1)(2p+l+2)\right),\nonumber\\
\int\Psi^\dagger\tilde r^2\Psi=\pi\frac{(l+p)!}{p!}\left(\left((\mathcal E+m)^2+k^2\right)(2p+l+1)+2B|e|(p+l)(2p+l)\right),\nonumber\\
\int\Psi^\dagger\tilde r^2\Psi=\pi\frac{(l+p)!}{p!}\left(\left((\mathcal E+m)^2+k^2\right)(2p+l+1)+2B|e|(p+1)(2p+l+2)\right),\nonumber\\
\int\Psi^\dagger\tilde r^2\Psi=\pi\frac{(l+p)!}{p!}\left(\left((\mathcal E+m)^2+k^2\right)(2p+l+1)+2B|e|p(2p+l)\right).\nonumber
\end{eqnarray}
By noting that $\mathcal E_L^2+\mathcal E_Z^2$ is $2B|e|$ times resp. $p+l+1$, $p+l$, $p+1$ and $p$ and rearranging, one gets
\begin{eqnarray}
\int\Psi^\dagger\tilde r^2\Psi=\pi\frac{(l+p)!}{p!}\left(2\mathcal E(\mathcal E+m)(2p+l+1)+\mathcal E_L^2+\mathcal E_Z^2\right),\nonumber\\
\int\Psi^\dagger\tilde r^2\Psi=\pi\frac{(l+p)!}{p!}\left(2\mathcal E(\mathcal E+m)(2p+l+1)-\mathcal E_L^2-\mathcal E_Z^2\right),\nonumber\\
\int\Psi^\dagger\tilde r^2\Psi=\pi\frac{(l+p)!}{p!}\left(2\mathcal E(\mathcal E+m)(2p+l+1)+\mathcal E_L^2+\mathcal E_Z^2\right),\nonumber\\
\int\Psi^\dagger\tilde r^2\Psi=\pi\frac{(l+p)!}{p!}\left(2\mathcal E(\mathcal E+m)(2p+l+1)-\mathcal E_L^2-\mathcal E_Z^2\right).\nonumber
\end{eqnarray}
Dividing by $\int\Psi^\dagger\Psi={\textstyle \int j_0}=2\pi\mathcal E(\mathcal E+m)\frac{(p+l)!}{p!}$ and adding the canonical angular momentum, resp. $l+\frac 12,\;l-\frac 12,\;-l+\frac 12$ and $-l-\frac 12$, one gets for the gauge covariant angular momentum
\begin{eqnarray}
&\mathcal J_z=&2p+2l+\frac 32+\frac{\mathcal E^2_L+\mathcal E^2_Z}{2\mathcal E(\mathcal E+m)},\nonumber\\
&\mathcal J_z=&2p+2l+\frac 12-\frac{\mathcal E^2_L+\mathcal E^2_Z}{2\mathcal E(\mathcal E+m)},\nonumber\\
&\mathcal J_z=&2p+\frac 32+\frac{\mathcal E^2_L+\mathcal E^2_Z}{2\mathcal E(\mathcal E+m)},\nonumber\\
&\mathcal J_z=&2p+\frac 12-\frac{\mathcal E^2_L+\mathcal E^2_Z}{2\mathcal E(\mathcal E+m)}.\nonumber
\end{eqnarray}
\section*{Computation of the magnetic moment}
Using the explicit form of the azimuthal Dirac matrix, it is easy to see that only the crossterms between the main and spin-orbit parts contribute to the azimuthal current and these can be computed to be
\begin{eqnarray}
&j_\phi=&2\sqrt 2\tilde r^{2l+1}e^{-\tilde r^2}L_p^l(\tilde r^2)L_p^{l+1}(\tilde r^2)(\mathcal E+m)\sqrt{B|e|},\nonumber\\
&j_\phi=&2\sqrt 2(p+l)\tilde r^{2l-1}e^{-\tilde r^2}L_p^l(\tilde r^2)L_p^{l-1}(\tilde r^2)(\mathcal E+m)\sqrt{B|e|},\nonumber\\
&j_\phi=&-2\sqrt 2(p+1)\tilde r^{2l-1}e^{-\tilde r^2}L_p^l(\tilde r^2)L_{p+1}^{l-1}(\tilde r^2)(\mathcal E+m)\sqrt{B|e|},\nonumber\\
&j_\phi=&-2\sqrt 2\tilde r^{2l+1}e^{-\tilde r^2}L_p^l(\tilde r^2)L_{p-1}^{l+1}(\tilde r^2)(\mathcal E+m)\sqrt{B|e|}.\nonumber
\end{eqnarray}
Now $M_z=\int\frac e2 rj_\phi \tilde r d\phi d\tilde r=\sqrt{\frac 2{B|e|}}\int\frac e2 \tilde rj_\phi r d\phi d\tilde r$. Substituting the explicit currents into this integral, using $x=\tilde r^2$ and performing the angular integration yields
\begin{eqnarray}
M_z=-2\pi(\mathcal E+m)|e|\int_0^\infty x^{l+1}L_p^l(x)L_p^{l+1}(x) e^{-x}dx,\nonumber\\
M_z=-2\pi(\mathcal E+m)|e|(p+l)\int_0^\infty x^lL_p^l(x)L_p^{l-1}(x) e^{-x}dx,\nonumber\\
M_z=2\pi(\mathcal E+m)|e|(p+1)\int_0^\infty x^lL_p^l(x)L_{p+1}^{l-1}(x) e^{-x}dx,\nonumber\\
M_z=2\pi(\mathcal E+m)|e|\int_0^\infty x^{l+1}L_p^l(x)L_{p-1}^{l+1}(x) e^{-x}dx.\nonumber
\end{eqnarray}
Now one can again use $L_p^l(x)=L_p^{l+1}(x)-L_{p-1}^{l+1}(x)$ and the orthogonality relation of associated Laguerre polynomials to evaluate these integrals
\begin{eqnarray}
M_z=-2\pi(\mathcal E+m)|e|\frac{(l+p+1)!}{p!}=-\frac{\textstyle{\int}j_0}{\mathcal E}\frac{\mathcal E_L^2+\mathcal E_Z^2}{2B},\nonumber\\
M_z=-2\pi(\mathcal E+m)|e|(p+l)\frac{(p+l)!}{p!}=-\frac{\textstyle{\int}j_0}{\mathcal E}\frac{\mathcal E_L^2+\mathcal E_Z^2}{2B},\nonumber\\
M_z=-2\pi(\mathcal E+m)|e|(p+1)\frac{(l+p)!}{p!}=-\frac{\textstyle{\int}j_0}{\mathcal E}\frac{\mathcal E_L^2+\mathcal E_Z^2}{2B},\nonumber\\
M_z=-2\pi(\mathcal E+m)|e|\frac{(l+p)!}{(p-1)!}=-\frac{\textstyle{\int}j_0}{\mathcal E}\frac{\mathcal E_L^2+\mathcal E_Z^2}{2B}.\nonumber
\end{eqnarray}
\section*{Commutator identities for the gauge covariant angular momentum operators}
For this section we write the angular omentum operators in antisymmetric tensor form. The gauge covariant angular momentum can be split in a spin and an orbital part like
$$
\mathcal J_{\mu\nu}=\mathcal L_{\mu\nu}+\frac i2\sigma_{\mu\nu},\qquad\mbox{with } \sigma_{\mu\nu }=\frac 12[\gamma_\mu,\gamma_\nu]\mbox{ and }\mathcal L_{\mu\nu}=x_{[\mu}P_{\nu]}\equiv x_\mu P_\nu-x_\nu P_\mu.
$$
Because $\mathcal L_{\mu\nu}$ contains no Dirac matrices, one obviously has $[\mathcal L_{\mu\nu},\sigma_{\rho\sigma}]=0$, so $[\mathcal J_{\mu\nu},\mathcal J_{\rho\sigma}]=[\mathcal L_{\mu\nu},\mathcal L_{\rho\sigma}]-\frac 14[\sigma_{\mu\nu},\sigma_{\rho\sigma}]$. Writing out the commutator for the $\sigma$-tensor gives
$$
[\sigma_{\mu\nu},\sigma_{\rho\sigma}]=\frac 14\left((\gamma_\mu\gamma_\nu-\gamma_\nu\gamma_\mu)(\gamma_\rho\gamma_\sigma-\gamma_\sigma\gamma_\rho)-(\gamma_\rho\gamma_\sigma-\gamma_\sigma\gamma_\rho)(\gamma_\mu\gamma_\nu-\gamma_\nu\gamma_\mu)\right).
$$
One can check that if all four indices are different, this commutator is zero. If two indices are the same one can eliminate the identical Dirac matrices and obtain after some algebra
$$
[\sigma_{\mu\nu},\sigma_{\rho\sigma}]=2(-\eta_{\mu\rho}\sigma_{\nu\sigma}+\eta_{\mu\sigma}\sigma_{\nu\rho}+\eta_{\nu\rho}\sigma_{\mu\sigma}-\eta_{\nu\sigma}\sigma_{\mu\rho} ).
$$
For the orbital part, we need the commutation relations of the gauge covariant momentum $P_\mu=i\partial_\mu-eA_\mu$. It is easy to check that
$$
[ix_{[\mu}\partial_{\nu]},ix_{[\rho}\partial_{\sigma]}]=-\eta_{\mu\rho}x_{[\nu}\partial_{\sigma]}+\eta_{\mu\sigma}x_{[\nu}\partial_{\rho]}
+\eta_{\nu\rho}x_{[\mu}\partial_{\sigma]}-\eta_{\nu\sigma}x_{[\mu}\partial_{\rho]}.
$$
To get the commutators for the gauge coveriant orbital angular momenta, we need to add $[ix_{[\mu}\partial_{\nu]},-ex_{[\rho}A_{\sigma]}]+[-ex_{[\mu}A_{\nu]},ix_{[\rho}\partial_{\sigma]}]=[ix_{[\mu}\partial_{\nu]},-ex_{[\rho}A_{\sigma]}]-[ix_{[\rho}\partial_{\sigma]},-ex_{[\mu}A_{\nu]}]$ (the vector potentias commute with each other). Using that both terms are the same up to the index swap $\mu\leftrightarrow\rho$, $\nu\leftrightarrow\sigma$ and using $\partial_\mu A_\rho-(\mu\leftrightarrow\rho)=F_{\mu\rho}$ these terms can be evaluated to be
\begin{multline}
[ix_{[\mu}\partial_{\nu]},-ex_{[\rho}A_{\sigma]}]+[-ex_{[\mu}A_{\nu]},ix_{[\rho}\partial_{\sigma]}]=-ie(\eta_{\mu\rho}x_{[\nu}A_{\sigma]}-\eta_{\nu\rho}x_{[\mu}A_{\sigma]}-\eta_{\mu\sigma} x_{[\nu}A_{\rho]}+\eta_{\nu\sigma}x_{[\mu}A_{\rho]})\\
-ie(x_\mu x_\rho F_{\nu\sigma}-x_\mu x_\sigma F_{\nu\rho}-x_\nu x_\rho F_{\mu\sigma}+x_\nu x_\sigma F_{\mu\rho}).\nonumber
\end{multline}
Using $-x_{[\mu}\partial_{\rho]}-ie x_{[\mu}A_{\rho]}=ix_{[\mu}P_{\rho]}$, and putting things together gives
\begin{eqnarray}
-i[\mathcal L_{\mu\nu},\mathcal L_{\rho\sigma}]=\eta_{\mu\rho}\mathcal L_{\nu\sigma}-\eta _{\mu\sigma}\mathcal L_{\nu\rho}-\eta_{\nu\rho}\mathcal L_{\mu\sigma}+\eta_{\nu\sigma}\mathcal L_{\mu\rho}+e(x_\mu x_\rho F_{\nu\sigma}-x_\mu x_\sigma F_{\nu\rho}-x_\nu x_\rho F_{\mu\sigma}+x_\nu x_\sigma F_{\mu\rho}),\nonumber\\
-i[\mathcal J_{\mu\nu},\mathcal J_{\rho\sigma}]=\eta_{\mu\rho}\mathcal J_{\nu\sigma}-\eta _{\mu\sigma}\mathcal J_{\nu\rho}-\eta_{\nu\rho}\mathcal J_{\mu\sigma}+\eta_{\nu\sigma}\mathcal J_{\mu\rho}+e(x_\mu x_\rho F_{\nu\sigma}-x_\mu x_\sigma F_{\nu\rho}-x_\nu x_\rho F_{\mu\sigma}+x_\nu x_\sigma F_{\mu\rho}).\nonumber
\end{eqnarray}
Then using $\mathcal J_x=\mathcal J_{23}$, $\mathcal J_y=\mathcal J_{31}$ and $\mathcal J_z=\mathcal J_{12}$, one gets 
$$
[\mathcal J_j,\mathcal J_k]=-i\epsilon_{jkl}(\mathcal J_l-x_l\mathbf{x\cdot B}),
$$
For the commutator $[\cancel P-m,J_{\mu\nu}]$, one can first note that $m$ commutes with any operator. Again using $\mathcal J_{\mu\nu}=\mathcal L_{\mu\nu}+\frac i2\sigma_{\mu\nu}$, one can compute the commutators of the spin and orbital parts seperately using $[P_\mu,P_\nu]=-ie F_{\mu\nu}$:
\begin{eqnarray}
[\cancel P,\sigma_{\mu\nu}]=\frac 12P^\lambda\left[ \gamma_\lambda,[\gamma_\mu,\gamma_\nu]\right]=2P^\lambda\eta_{\lambda[\mu}\gamma_{\nu]}=2P_{[\mu}\gamma_{\nu]}=-2 \gamma_{[\mu}P_{\nu]},\nonumber\\
\left[\gamma^\lambda P_\lambda,x_{[\mu}P_{\nu]}\right]=i\gamma^{\lambda}\eta_{\lambda[\mu}P_{\nu]}+iex_{[\mu}F_{\nu]\lambda}\gamma^{\lambda}=i\gamma_{[\mu}P_{\nu]}+iex_{[\mu}F_{\nu]\lambda}\gamma^{\lambda},\nonumber\\
\left[\gamma^\lambda P_\lambda,x_{[\mu}P_{\nu]}+\frac i2\sigma_{\mu\nu}\right]=[\cancel P,J_{\mu\nu}]=iex_{[\mu}F_{\nu]\lambda}\gamma^{\lambda}.\nonumber
\end{eqnarray}
\end{document}